\documentclass[preprintnumbers,amsmath,amssymb,showpacs,twocolumn]{revtex4}

\usepackage{graphicx}
\usepackage{dcolumn}
\usepackage{bm}

\begin{document}

\title{Probabilistic Dense Coding Using Non-Maximally Entangled
Three-Particle State}

\author{ZHANG Guo-Hua, YAN Feng-Li}
\thanks{Corresponding author. Email address: flyan@mail.hebtu.edu.cn}
 \affiliation {College of Physics Science and Information
Engineering, Hebei Normal University, Shijiazhuang 050016, China;\\
Hebei Advanced Thin Films Laboratory,  Shijiazhuang 050016, China}

\date{\today}

\begin{abstract}
We present a scheme of probabilistic dense coding via a quantum
channel of non-maximally entangled three-particle state. The quantum
dense coding will be succeeded  with a certain probability if the
sender introduces an auxiliary particle and performs a collective
unitary transformation. Furthermore, the average information
transmitted in this scheme is calculated.
\end{abstract}


\maketitle

One of the essential features of quantum information is its capacity
for entanglement.  The present-day entanglement theory has its roots
in the key discoveries: quantum teleportation,$^{[1]}$ quantum
cryptography with Bell theorem,$^{[2]}$ and quantum dense
coding.$^{[3]}$  Entanglement has also played important role in
development of quantum computation and quantum communication.
$^{[4-7]}$

Holevo has shown that one qubit can carry at most only one bit of
classical information. $^{[8]}$  In 1992, Bennett and Wiesner
discovered a fundamental primitive, quantum dense coding, $^{[3]}$
which allows to communicate two classical bits by sending one a
priori entangled qubit.  Quantum dense coding is one of many
surprising applications of quantum entanglement. In 1996, quantum
dense coding was experimentally demonstrated with polarization
entangled photons for the case of discrete variables by Mattle et
al. $^{[9]}$ Recent years,     some schemes for quantum dense coding
using multi-particle entangled states via local
measurements,$^{[10]}$ GHZ state,$^{[11]}$ non-symmetric quantum
channel $^{[12]}$ were proposed. Liu et al.$^{[13]}$ presented the
general scheme for dense coding between multi-parties using a
high-dimensional state. All these cases deal with maximally
entangled states.

Recently, Hao et al.$^{[14]}$ gave  a probabilistic dense coding
scheme using the two-qubit pure state
$|\phi\rangle=a|00\rangle+b|11\rangle$. A general probabilistic
dense coding scheme was put forward by Wang et al $^{[15]}$.

In this Letter,  we suggest a scheme of probabilistic dense coding
via a quantum channel of non-maximally entangled three-particle
state. The average information transmitted in the scheme is
calculated. Furthermore, the scheme is generalized to $d$-level
 $(d>3)$ for three parties.

We first consider the  dense coding between three-parties (Alice,
Bob and Charlie) via a  maximally entangled three-particle state.
Suppose Alice and Bob are the senders, Charlie is the receiver,  a
maximally entangled three-particle state
\begin{equation} |\psi_{00}\rangle_{ABC}=\frac {1}{\sqrt
3}(|000\rangle+|111\rangle+|222\rangle)_{ABC}
\end{equation}
is shared by them, and the three particles $A$, $B$ and $C$ are held
by Alice, Bob and Charlie, respectively.

 Let us introduce the nine  single-particle
operations  as follows:
\begin{equation}
\begin{array}{ll}
U_{00}=\left (\begin{array}{ccc} 1&0&0\\
0&1&0\\
0&0&1\end{array}\right), &U_{01}=\left (\begin{array}{ccc} 1&0&0\\
0&e^{2\pi i/3}&0\\
0&0&e^{4\pi i/3}\end{array}\right),\\
U_{02}=\left (\begin{array}{ccc} 1&0&0\\
0&e^{4\pi i/3}&0\\
0&0&e^{2\pi i/3}\end{array}\right), &U_{10}=\left (\begin{array}{ccc} 0&0&1\\
1&0&0\\
0&1&0\end{array}\right),\\
U_{11}=\left (\begin{array}{ccc} 0&0&e^{4\pi i/3}\\
1&0&0\\
0&e^{2\pi i/3}&0\end{array}\right), &U_{12}=\left (\begin{array}{ccc} 0&0&e^{2\pi i/3}\\
1&0&0\\
0&e^{4\pi i/3}&0\end{array}\right),\\
U_{20}=\left (\begin{array}{ccc} 0&1&0\\
0&0&1\\
1&0&0\end{array}\right), &U_{21}=\left (\begin{array}{ccc} 0&e^{2\pi i/3}&0\\
0&0&e^{4\pi i/3}\\
1&0&0\end{array}\right),\\
U_{22}=\left (\begin{array}{ccc} 0&e^{4\pi i/3}&0\\
0&0&e^{2\pi i/3}\\
1&0&0\end{array}\right).\end{array}\end{equation}

It is easy to prove that
\begin{eqnarray}
&&U_{00}(A)\otimes U_{00}(B)|\psi_{00}\rangle_{ABC}\\\nonumber
 &=&\frac
{1}{\sqrt
3}(|000\rangle+|111\rangle+|222\rangle)_{ABC}\equiv|\psi^0_{00}\rangle_{ABC},\\\nonumber
&&U_{00}(A)\otimes U_{10}(B)|\psi_{00}\rangle_{ABC}\\\nonumber
 &=&\frac
{1}{\sqrt
3}(|010\rangle+|121\rangle+|202\rangle)_{ABC}\equiv|\psi^0_{01}\rangle_{ABC},\\\nonumber
&&U_{00}(A)\otimes U_{20}(B)|\psi_{00}\rangle_{ABC}\\\nonumber &
=&\frac {1}{\sqrt
3}(|020\rangle+|101\rangle+|212\rangle)_{ABC}\equiv|\psi^0_{02}\rangle_{ABC},\\\nonumber
&&U_{01}(A)\otimes U_{00}(B)|\psi_{00}\rangle_{ABC}\\\nonumber
&=&\frac {1}{\sqrt 3}(|000\rangle+e^{2\pi i/3}|111\rangle+e^{4\pi
i/3}|222\rangle)_{ABC}\equiv|\psi^0_{10}\rangle_{ABC},\\\nonumber
&&U_{01}(A)\otimes U_{10}(B)|\psi_{00}\rangle_{ABC}\\\nonumber
 &=&\frac
{1}{\sqrt 3}(|010\rangle+e^{2\pi i/3}|121\rangle+e^{4\pi
i/3}|202\rangle)_{ABC}\equiv|\psi^0_{11}\rangle_{ABC},\\\nonumber
&&\cdots,\\\nonumber
 &&U_{22}(A)\otimes
U_{20}(B)|\psi_{00}\rangle_{ABC}\\\nonumber
 & =&\frac {1}{\sqrt
3}(|220\rangle+e^{4\pi i/3}|001\rangle+e^{2\pi
i/3}|112\rangle)_{ABC}\equiv|\psi^2_{22}\rangle_{ABC}\end{eqnarray}
and
$\{|\psi^0_{00}\rangle,|\psi^0_{01}\rangle,|\psi^0_{02}\rangle,|\psi^0_{10}\rangle,|\psi^0_{11}\rangle,\cdots,|\psi^2_{22}\rangle
\}$ is  a  basis of the Hilbert space of the particles $A$, $B$ and
$C$.

Alice performs one of the nine unitary transformations stated in
Eq.(2) on particle $A$, Bob operates one of  three unitary
transformations $U_{00}, U_{10} , U_{20}$, on particle $B$. Then
they send their particles $A$ and $B$ to the receiver Charlie. After
receiving the particles $A$ and $B$,  Charlie takes only one
measurement in the basis
$\{|\psi^0_{00}\rangle,|\psi^0_{01}\rangle,|\psi^0_{02}\rangle,|\psi^0_{10}\rangle,|\psi^0_{11}\rangle,\cdots,|\psi^2_{22}\rangle
\}$, and she will know what operation Alice and Bob have carried
out, that is, what messages are that Alice and Bob have encoded in
the quantum state. Then  Charlie can obtain ${\rm log}_2 27$ bits of
information through only one measurement. So the dense coding is
realized successfully.

In the following we will discuss a scheme of probabilistic dense
coding between three parties via a  non-maximally entangled
three-particle state. We suppose that Alice, Bob and Charlie share a
non-maximally entangled three-particle state
\begin{equation}
|\varphi\rangle_{ABC}=(x_0|000\rangle+x_1|111\rangle+x_2|222\rangle)_{ABC},
\end{equation}
where $x_0, x_1, x_2$  are real numbers, and
$|x_0|^2+|x_1|^2+|x_2|^2=1$. Without loss of generality, we can
suppose that $|x_0|\leq |x_1|\leq |x_2|$.

   The scheme of probabilistic dense coding is composed of four steps.

   Firstly, Alice introduces an auxiliary three-level particle $a$ in the quantum state $|0\rangle_a$.
   Then she performs a unitary transformation  $U_{sim}$ on her particle A and auxiliary particle $a$ under the basis
    $\{|00\rangle_{Aa}, |01\rangle_{Aa}, |02\rangle_{Aa},|10\rangle_{Aa},|11\rangle_{Aa},|12\rangle_{Aa},|20\rangle_{Aa},$
    $|21\rangle_{Aa},|22\rangle_{Aa}\}$:
\begin{equation}
U_{sim}=\left (\begin{array}{ccccccccc} 1&0&0&0&0&0&0&0&0\\
0&1&0&0&0&0&0&0&0\\
0&0&1&0&0&0&0&0&0\\
0&0&0&m_{01}&m&0&0&0&0\\
0&0&0&m&-m_{01}&0&0&0&0\\
0&0&0&0&0&m_{02}&0&M&-N\\
0&0&0&0&0&0&m_{02}&N&M\\
0&0&0&0&0&M&N&-m_{02}&0\\
0&0&0&0&0&-N&M&0&-m_{02}\\
\end{array}\right),
\end{equation}
where  $M=\sqrt {(x^2_2-x^2_1)/x_2^2},N=\sqrt {(x^2_1-x^2_0)/x_2^2}
,m=\sqrt {1-x^2_0/x_1^2}, m_{01}=x_0/x_1, m_{02}=x_0/x_2$. The
collective unitary operator $U_{sim}\otimes I_{BC}$ (where $I_{BC}$
is a $9\times9$ identity matrix) transforms the state
$|\varphi\rangle_{ABC}\otimes |0\rangle_a$ into
\begin{eqnarray}
&&|\varphi'\rangle_{ABCa}\\\nonumber &=&U_{sim}\otimes
I_{BC}|\varphi\rangle_{ABC}\otimes |0\rangle_a\\\nonumber &=&\sqrt 3
x_0\frac {1}{\sqrt
3}(|000\rangle+|111\rangle+|222\rangle)_{ABC}|0\rangle_a\\\nonumber
&&+\sqrt 2 \sqrt {x_1^2-x_0^2}\frac {1}{\sqrt
2}(|111\rangle+|222\rangle)_{ABC}|1\rangle_a\\\nonumber &&+\sqrt
{x_2^2-x_1^2}|222\rangle_{ABC}|2\rangle_a\\\nonumber &\equiv&\sqrt 3
x_0|\varphi_{00}\rangle_{ABC}|0\rangle_a+\sqrt 2 \sqrt
{x_1^2-x_0^2}|\varphi'_{00}\rangle_{ABC}|1\rangle_a\\\nonumber
&&+\sqrt {x_2^2-x_1^2}|\varphi''_{00}\rangle_{ABC}|2\rangle_a.
\end{eqnarray}
Then Alice makes a measurement on the auxiliary particle $a$ and
tells Bob and Charlie her measurement result via  a classical
channel. If she gets the result $|0\rangle_a$, she ensures that the
three particles $A$, $B$ and $C$ are in the maximally entangled
three-particle state $|\varphi_{00}\rangle=\frac {1}{\sqrt
3}(|000\rangle+|111\rangle+|222\rangle)$, the probability of
obtaining $|0\rangle_a$ is $3x_0^2$ according to Eq.(6); if  the
result $|1\rangle_a$ is obtained, she ensures that the three
particles $A$, $B$ and $C$ are in the state
$|\varphi'_{00}\rangle=\frac {1}{\sqrt 2}(|111\rangle+|222\rangle)$,
and the probability of getting this result is $2(x_1^2-x_0^2)$; if
she gets the result $|2\rangle_a$, she knows that the three
particles are in the product state
$|\varphi''_{00}\rangle=|222\rangle$, and the probability of
obtaining this result is $x_2^2-x_1^2$.

Secondly, Alice and Bob encode classical information by performing
the unitary transformations on their particle $A$ and $B$
respectively.

 If the
three particles $A$, $B$ and $C$ are in the state
$|\varphi_{00}\rangle=\frac {1}{\sqrt
3}(|000\rangle+|111\rangle+|222\rangle)$, one  uses the dense coding
protocol stated above. Here we do not recite any more.

If the three particles $A$, $B$ and $C$ are in the state
$|\varphi'_{00}\rangle=\frac {1}{\sqrt 2}(|111\rangle+|222\rangle)$,
Alice encodes  her message by performing  one of six single-particle
operations
\begin{equation}
\begin{array}{ll}
U'_{00}=\left (\begin{array}{ccc} 1&0&0\\
0&1&0\\
0&0&1\end{array}\right), &U'_{01}=\left (\begin{array}{ccc} 1&0&0\\
0&1&0\\
0&0&-1\end{array}\right),\\
U'_{10}=\left (\begin{array}{ccc} 0&0&1\\
1&0&0\\
0&1&0\end{array}\right), &U'_{11}=\left (\begin{array}{ccc} 0&0&-1\\
1&0&0\\
0&1&0\end{array}\right),\\
U'_{20}=\left (\begin{array}{ccc} 0&1&0\\
0&0&1\\
1&0&0\end{array}\right), &U'_{21}=\left (\begin{array}{ccc} 0&1&0\\
0&0&-1\\
1&0&0\end{array}\right)\end{array}\end{equation}
on particle $A$.
However, Bob encodes his message only by operating one of three
single-particle operations $U'_{00}$, $U'_{10}$, $U'_{20}$ on
particle $B$. Through simple calculation, we can prove that
\begin{eqnarray}
&&U'_{00}(A)\otimes U'_{00}(B)|\varphi'_{00}\rangle_{ABC}\\\nonumber
 &=&\frac
{1}{\sqrt
2}(|111\rangle+|222\rangle)_{ABC}\equiv|\varphi'^0_{00}\rangle_{ABC},\\\nonumber
&&U'_{00}(A)\otimes U'_{10}(B)|\varphi'_{00}\rangle_{ABC}\\\nonumber
 &=&\frac
{1}{\sqrt
2}(|121\rangle+|202\rangle)_{ABC}\equiv|\varphi'^1_{00}\rangle_{ABC},\\\nonumber
&&U'_{00}(A)\otimes U'_{20}(B)|\varphi'_{00}\rangle_{ABC}\\\nonumber
& =&\frac {1}{\sqrt
2}(|101\rangle+|212\rangle)_{ABC}\equiv|\varphi'^2_{00}\rangle_{ABC},\\\nonumber
&&U'_{01}(A)\otimes U'_{00}(B)|\varphi'_{00}\rangle_{ABC}\\\nonumber
&=&\frac {1}{\sqrt
2}(|111\rangle-|222\rangle)_{ABC}\equiv|\varphi'^0_{01}\rangle_{ABC},\\\nonumber
&&\cdots,\\\nonumber &&U'_{21}(A)\otimes
U'_{20}(B)|\varphi'_{00}\rangle_{ABC}\\\nonumber
 & =&\frac {1}{\sqrt
2}(|001\rangle-|112\rangle)_{ABC}\equiv|\varphi'^2_{21}\rangle_{ABC}.\end{eqnarray}
It is easy to see that the states in the set
$\{|\varphi'^k_{mn}\rangle, m,k=0,1,2; n=0,1\}$ are orthogonal each
other.

If the three particles $A$, $B$ and $C$ are in the product state
$|\varphi''_{00}\rangle=|222\rangle$, Alice and Bob can encode their
classical information by performing  one of three single-particle
operations on particles $A$ and $B$ independently:
\begin{equation}
\begin{array}{ll}
U''_{00}=\left (\begin{array}{ccc} 1&0&0\\
0&1&0\\
0&0&1\end{array}\right), &U''_{10}=\left (\begin{array}{ccc} 0&0&1\\
1&0&0\\
0&1&0\end{array}\right),\\
U''_{20}=\left (\begin{array}{ccc} 0&1&0\\
0&0&1\\
1&0&0\end{array}\right).\end{array}\end{equation} The state
$|\varphi''_{00}\rangle$ will be transformed into the corresponding
state respectively:
\begin{eqnarray}
\\\nonumber&&U''_{00}(A)\otimes
U''_{00}(B)|\varphi''_{00}\rangle_{ABC}=|222\rangle_{ABC}\equiv|\varphi''_{00}\rangle_{ABC},\\\nonumber
&&U''_{00}(A)\otimes U''_{10}(B)|\varphi''_{00}\rangle_{ABC}
 =|202\rangle_{ABC}\equiv|\varphi''_{01}\rangle_{ABC},\\\nonumber
&&U''_{00}(A)\otimes U''_{20}(B)|\varphi''_{00}\rangle_{ABC}
=|212\rangle_{ABC}\equiv|\varphi''_{02}\rangle_{ABC},\\\nonumber
&&U''_{10}(A)\otimes
U''_{00}(B)|\varphi''_{00}\rangle_{ABC}=|022\rangle_{ABC}\equiv|\varphi''_{10}\rangle_{ABC},\\\nonumber
&&\cdots,\\\nonumber &&U''_{20}(A)\otimes
U''_{20}(B)|\varphi''_{00}\rangle_{ABC}=|112\rangle_{ABC}\equiv|\varphi''_{22}\rangle_{ABC}.\end{eqnarray}
Evidently, the states in the set $\{|\varphi''_{mn}\rangle,
m,n=0,1,2 \}$ are orthogonal each other.

Thirdly, Alice and Bob send their particles $A$ and $B$
independently to Charlie.

Finally, After Charlie receives particle $A$ and $B$, she takes only
one measurement on the three particle $A$, $B$ and $C$. The
measurement basis is determined by  Alice's measurement result.
According to Charlie's measurement result, Charlie will know what
operators Alice and Bob have carried out, i.e. he can obtain the
classical information that Alice and Bob have encoded.

Apparently, the average information transmitted in this procedure is
\begin{equation}
I_{aver}=3x_0^2{\rm log}_2 27+2(x_1^2-x_0^2){\rm log}_2
18+(x_2^2-x_1^2){\rm log}_2 9.
\end{equation}
In fact, the above protocol needs $2{\rm log}_2 3$ bits of classical
information for Alice to tell Bob and Charlie her measurement result
on the auxiliary particle. Obviously, when $x_0=x_1=x_2=\frac
{1}{\sqrt 3} $, the three particles $A$, $B$ and $C$ is in  the
maximally entangled three-particle state, and the success
probability of dense coding is one. The average information
transmitted is ${\rm log}_2 27$ bits.

Now we would like to  generalize the above protocol   to $d$-level
 for three parties.  Suppose that Alice, Bob and Charlie share a non-maximally
entangled three-particle state
\begin{equation}
|\varphi\rangle_{ABC}=(x_0|000\rangle+x_1|111\rangle+\cdots+x_{d-1}|d-1d-1d-1\rangle)_{ABC},
\end{equation}
where $x_0,x_1,\cdots,x_{d-1}$  are real numbers and satisfy
$|x_0|\leq |x_1|\leq \cdots \leq |x_{d-1}|$.

The scheme of the probabilistic dense coding can be accomplished by
four steps.

(1) Alice introduces an auxiliary $d$-level particle in the quantum
state $|0\rangle_a$. Then she performs a proper unitary
transformation on her particle $A$ and the auxiliary particle. The
collective unitary transformation $U_{sim}\otimes I_{BC}$ (where
$I_{BC}$ is a $d^2\times d^2$ identity matrix) transforms the state
$|\varphi\rangle_{ABC}\otimes |0\rangle_a$ into the state
\begin{eqnarray}
&&|\varphi\rangle_{ABCa}\\\nonumber
 &=&
x_0(|000\rangle+|111\rangle+\cdots+|d-1d-1d-1\rangle)_{ABC}|0\rangle_a\\\nonumber
&&+\sqrt
{x_1^2-x_0^2}(|111\rangle+\cdots+|d-1d-1d-1\rangle)_{ABC}|1\rangle_a\\\nonumber
&&+\cdots+\sqrt
{x_{d-1}^2-x_{d-2}^2}|d-1d-1d-1\rangle_{ABC}|d-1\rangle_a.
\end{eqnarray}
After that  Alice performs a measurement on the auxiliary particle.
The resulting state of the particles $A$, $B$ and $C$ will be
respectively

 $\frac {1}{\sqrt
d}(|000\rangle+|111\rangle+\cdots+|d-1d-1d-1\rangle),$

 $\frac
{1}{\sqrt {d-1}}(|111\rangle+\cdots+|d-1d-1d-1\rangle),$

 $\cdots,$

 $|d-1d-1d-1\rangle$.\\
The probability of obtaining each resulting state is $dx_0^2$,
$(d-1)(x_1^2-x_0^2), \cdots, (x_{d-1}^2-x_{d-2}^2)$, respectively.

(2) Alice tells Bob and Charlie her measurement result, then Alice
and Bob encode classical information by making a unitary
transformation on  particle $A$ and $B$ respectively.

(3) Alice and Bob send  particle $A$ and $B$ to Charlie
respectively.

(4) After Charlie receives the particles $A$ and $B$, she takes a
measurement in the basis determined by Alice's measurement result.
According to his measurement result, Charlie can obtain the
classical information that Alice and Bob have encoded. The average
information transmitted
 is
\begin{eqnarray}
&&I_{aver}\\\nonumber &=&dx_0^2{\rm log}_2
d^3+(d-1)(x_1^2-x_0^2){\rm log}_2 d^2(d-1)\\\nonumber
&&+(d-2)(x_2^2-x_1^2){\rm log}_2 d^2(d-2)+\cdots\\\nonumber
&&+(x_{d-1}^2-x_{d-2}^2){\rm log}_2 d^2.
\end{eqnarray}

Obviously, this probabilistic dense coding scheme needs $2{\rm
log}_2 d$ bits of information to transmit Alice's measurement
results on the auxiliary particle to Bob and Charlie.

It is easy to see that if the non-zero coefficients in Eq.(12) are
totally equal, Alice does not need to introduce the auxiliary
particle and makes unitary transformation $U_{sim}$. The classical
information can be encoded directly by performing single-particle
operations on particle $A$ and $B$ respectively. In this case the
quantum state is called a deterministic quantum channel; otherwise,
 a probabilistic one if the coefficients
are not equal totally. Obviously, a non-maximally entangled
three-particle state is not equivalent to the probabilistic quantum
channel. For example, in the $3\otimes3\otimes3$-dimensional case,
the quantum state $\frac {1}{\sqrt 2}(|111\rangle+|222\rangle)$ is
not a maximally entangled three-particle state, but it is a
deterministic quantum channel. So  the first step in our protocol is
to extract a series of deterministic quantum  channels from a
probabilistic one.

In summary, we have presented a scheme of probabilistic dense coding
via a quantum channel of non-maximally entangled three-particle
state.  The average information transmitted in this scheme
 is explicitly given. We also generalize this scheme to the more
 general case.

{\noindent\bf Acknowledgments}\\[0.2cm]

 This work was supported by the National  Natural Science Foundation of
China under Grant No: 10671054, Hebei Natural Science Foundation of
China under Grant No: 07M006 and the Key Project of Science and
Technology Research of Education Ministry of China under Grant
No:207011.

\end{document}